\RequirePackage{fix-cm}
\documentclass[smallextended]{svjour3}
\usepackage{graphicx,amsmath,mathptmx,enumerate,latexsym,bm}
\usepackage[misc]{ifsym}
\usepackage[authoryear]{natbib}
\bibpunct{(}{)}{,}{a}{}{;}

\journalname{Mathematical Geoscience}

\begin{document}

\title{Uncoupling electrokinetic flow solutions}
\author{Kristopher L. Kuhlman \and Bwalya Malama}
\institute{K.L. Kuhlman \at Sandia National Laboratories\\
  Applied Systems Analysis and Research Department\\ Albuquerque, NM\\
  \email{klkuhlm@sandia.gov}
  \and
  B. Malama \at California Polytechnic State University, San Luis Obispo\\
  Natural Resources Management \& Environmental Sciences\\
  San Luis Obispo, CA\\
\email{bmalama@calpoly.edu}}

\date{}
\maketitle

\begin{abstract}
  The continuum-scale electrokinetic porous-media flow and excess charge redistribution equations are uncoupled using eigenvalue decomposition. The uncoupling results in a pair of independent diffusion equations for ``intermediate'' potentials subject to modified material properties and boundary conditions. The fluid pressure and electrostatic potential are then found by recombining the solutions to the two intermediate uncoupled problems in a matrix-vector multiply. Expressions for the material properties or source terms in the intermediate uncoupled problem may require extended precision or careful re-writing to avoid numerical cancellation, but the solutions themselves can be computed in typical double precision. The approach works with analytical or gridded numerical solutions and is illustrated through two examples. The solution for flow to a pumping well is manipulated to predict streaming potential and electroosmosis, and a periodic one-dimensional analytical solution is derived and used to predict electroosmosis and streaming potential in a laboratory flow cell subjected to low frequency alternating current and pressure excitation. The examples illustrate the utility of the eigenvalue decoupling approach, repurposing existing analytical solutions and leveraging simpler-to-derive solutions or numerical models for coupled physics.
\keywords{streaming potential \and electroosmosis \and electrokinetic \and eigenvalue \and coupled processes \and geophysics}
\end{abstract}

\section{Introduction}
Coupled physical phenomena exist at the intersection of hydrological and geophysical processes, and advanced solution methods are required to make general predictions. Radioactive waste disposal \citep{TSANG201231}, water resource management \citep{barthel16}, geothermal energy production \citep{baechler05}, and hydrogeophysics \citep{hinnell10} are examples of applications requiring holistic approaches to multiphysics. Although several alternative commercial \citep{li09} and research \citep{FiPy:2009,gaston09,liu13} software libraries exist to solve coupled physics using finite element or finite volume numerical methods, we illustrate an approach allowing adaptation of existing analytical or numerical methods when certain symmetries exist in the governing equations and boundary conditions. Coupled analytical solutions derived using this approach may serve to validate solutions obtained with traditional numerical methods.

While many exploration geophysical methods are sensitive to the presence or storage of water or solutes (e.g., electrical resistivity \citep{pollock12}, seismic \citep{hyndman94}, or induced polarization \protect \citep{ahmed2019multiscale}), in electrokinetics the coupling between hydrology and geophysics is especially explicit and direct \citep{revil13,revil15}. Direct causality exists between processes of flow and electric potentials for streaming potentials and electrokinetics. Streaming potentials are caused by the movement of water through a porous medium and electroosmosis is the movement water in low-permeability porous media due to applied electric fields.  Electrokinetics occur when electrolytes (e.g., water and ions) flow through a porous medium with a surface charge (e.g., quartz sand grains) \citep{pride94,bockris02}. Streaming potentials arise from the movement of water under an imposed pressure gradient, dragging ions with the water, creating a streaming current. Electroosmotic pressure arises from the movement of ions under an imposed electric field, dragging the water with the ions, creating an electroosmotic flux. Here, we account for both coupled electrokinetic processes occurring in a porous medium by decoupling the two governing equations and boundary conditions, solving for intermediate potentials, and recombining to get back the physical solution. The approach relies on transient effects in both of the governing equations, boundary conditions of the same type at each location in the electrical and hydrological problems, and equivalent anisotropy in the material properties of the electrical and hydrologic problems.  

Streaming potentials are useful at both the laboratory and field scales to characterize water movement using passive (i.e., self-potential without an applied current) voltage observations \citep{revil13,revil17a}. Streaming potentials have been used to derive material properties from field-scale pumping tests experiments \citep{revil2008hysteresis,malama09-confinedSP,malama09-unconfinedSP,malama2014theory,soueid2016joint}. Electroosmosis is utilized widely in microfluidics \citep{karniadakis05} and at the pore scale \citep{coelho96,gupta08} to move fluids through small pores. At the laboratory and field scale it has been used to consolidate soft clays \citep{banerjee80,banerjee84,lo91,alshawabkeh04} or mobilize contaminants  \citep{bruell1992electroosomotic,acar93,shapiro93,acar96,virkutyte02,bertolini2009electroosmotic}. Seismoelectric applications consider the electrokinetic response of a porous formation to seismic waves \citep{pride94,haines2006seismoelectric,revil15,peng2019effect}.

Predictions of streaming potentials and electroosmosis responses can be made using analytical solutions, but typically these solutions are one-way coupled (i.e., only explicitly accounting for one of the two conjugate electrokinetic effects). For example, typical streaming potential solutions use the existing solution for flow to a well (ignoring electroosmotic effects) as a source term in the electrical conduction equation \citep{malama09-unconfinedSP,malama09-confinedSP,malama2014theory}. A typical electroosmosis solution uses the existing solutions for electrical conduction around an electrode (ignoring streaming potential effects) as a source term in the flow equation \citep{banerjee80,shapiro93,reppert02}.  Few fully coupled porous-media scale electrokinetic solutions exist; \citet{pengra99} showed under simplified laboratory conditions independent measurement of electroosmotic pressure and streaming potentials can be used to estimate the permeability of porous materials. More general predictions of  fully coupled electrokinetics are usually done with finite-element or finite-volume multiphysics solution libraries like COMSOL or OpenFOAM \citep{probstein94,masliyah06,revil15}. 

Electrokinetic flow through porous media is presented here in a matrix formulation, with the two equations decoupled using an eigenvalue decomposition. The uncoupled solutions are two simpler diffusion equation solutions that can more readily be solved using analytical solutions or existing numerical models; solutions to these intermediate solutions are then recombined to solve the original electrokinetic problem. The approach should be applicable to coupling of constitutive laws including Darcy's law (porous media flow), Fourier's law (heat conduction), and Ohm's law (charge conduction) because the system of coupled equations can be represented as a symmetric system of equations, because these types of processes lead to the required types of symmetric equations. 

While an analogous uncoupling approach has been used in quantum mechanics for second-order inelastic scattering equations \citep{stechel78,light79} and in geophysics to decouple poroelastic and acoustic wave equations \citep{lo09}, the method has not previously been applied to uncouple electrokinetic phenomena. The authors were inspired for this approach by the eigenvalue approach to uncoupling a multilayered aquifer and aquitard flow system found in publications by Maas and Hemker \citep{hemker84JH,maas86,maas87b}. Following the analogy with the layered aquifer system, the coupled equations in matrix form could also be solved directly using special functions of matrix argument \citep{maas86}.

In Section~\ref{sec:PDES} we present the governing equations for electrokinetic flow and re-write them in dimensionless form. Section~\ref{sec:uncoupling} presents the uncoupling process for both the governing equations and boundary conditions. Section~\ref{sec:applications} shows two examples. The first example (Section~\ref{section:theis}) uses type-curve methods to represent electrokinetic flow to a well, and the second example (Section~\ref{sec:core}) predicts the response of a laboratory sample to periodic pressure and electrical excitation. Finally, we summarize the capabilities and limitations of the approach (Section~\ref{sec:limits}).

\section{Electrokinetic Governing Equations}\label{sec:PDES}
The decoupling approach relies on symmetry in the governing equations and boundary conditions.  We begin with expressions for the Darcy and electric current densities (i.e., fluxes) in terms of potential gradients (i.e., thermodynamic forces)
\begin{equation}
\begin{aligned}
  \label{eq:flux}
  \bm{j}_e &= -\sigma_0 \nabla \psi - L_{12} \nabla p \\
  \bm{j}_f &= -L_{21} \nabla \psi - \frac{k_0}{\mu} \nabla p
\end{aligned}
\end{equation}
where $\bm{j}_e$ and $\bm{j}_f$ are electrical $\mathrm{[A/m^2]}$ and Darcy (i.e., fluid-volume) $\mathrm{[m/sec]}$ current densities, $\sigma_0$ is bulk electrical conductivity $\mathrm{[S/m]}$, $k_0$ is porous medium intrinsic permeability $\mathrm{[m^2]}$, $\mu$ is pore water dynamic viscosity $\mathrm{[Pa \cdot sec]}$, \{$L_{12}$, $L_{21}$\} are porous medium electrokinetic coupling coefficients \{$\mathrm{[A/m \cdot Pa]}$, $\mathrm{[m^2/V\cdot sec]}$\}, $\psi$ is change in electrostatic potential $\mathrm{[V]}$ (in the quasi-static limit with no magnetic sources), and $p$ is change in liquid pressure $\mathrm{[Pa]}$. Both $p$ and $\psi$ are changes from an arbitrary initial state due to some forcing (e.g., pumping or recharge). The zero subscripts on $k$ and $\sigma$ differentiate them from similar quantities (i.e., $\bm{j}_{\mathrm{Darcy}}=-k/\mu \, \nabla p$ or $\bm{j}_{\mathrm{Ohm}}=-\sigma \, \nabla \psi$) that do not consider electrokinetic coupling effects \citep{pengra99}.

\citet{revil13} present the streaming potential coupling coefficient in terms of pore-scale electrokinetic quantities \citep{bockris02} as
\begin{equation}
  \label{eq:sp-zeta}
  L_{12}=\frac{\epsilon \zeta \sigma_0}{\mu \left( \sigma_f + F \sigma_s \right)}
\end{equation}
where $\epsilon$ is the pore-fluid dielectric constant $\mathrm{[F/m]}$, $\zeta$ is the zeta potential at the pore/fluid interface $\mathrm{[V]}$, $\sigma_f$ and $\sigma_s$ are the electrical conductivity of the individual fluid and solid components, and $F$ is the dimensionless formation factor, defined as the resistivity of the fluid-saturated rock normalized by the pore fluid resistivity.  In the geotechnical literature, $L_{21}$ is electroosmotic permeability ($k_e$) with a similar definition in terms of pore-scale quantities \citep{casagrande49,arnold73}. 

Mass and charge conservation expressions are conveniently and symmetrically written in terms of flux divergence as
\begin{align}
  \label{eq:conserv}
  -\nabla \cdot \bm{j}_e & = C^{\ast} \frac{\partial \psi}{\partial t}\\
  -\nabla\cdot \bm{j}_f & = nc \frac{\partial p}{\partial t},\nonumber
\end{align}
where $n$ is dimensionless porosity, $c$ is compressibility $\mathrm{[1/Pa]}$, and $C^{\ast}$ is specific capacitance $\mathrm{[C/(m^3 \cdot V)]}$, that is electrolyte charge flowing into a unit volume per unit change in potential \citep{corapcioglu91}.  Typically, in streaming potential problems $\nabla \cdot \bm{j}_e = 0$ \citep{revil13}, but without loss of generality we include a small transient capacitance term to maintain symmetry in the governing equations required by the uncoupling approach.

Substituting the fluxes \eqref{eq:flux} into the conservation equations \eqref{eq:conserv} leads to two coupled differential equations in terms of potentials,
\begin{align}
  \label{eq:pdes}
  C^{\ast} \frac{\partial \psi}{\partial t} &= \nabla \cdot \left(\sigma_0 \nabla \psi \right) + \nabla \cdot \left( L_{12} \nabla p \right) \\
  n c \frac{\partial p}{\partial t} &=  \nabla \cdot \left( L_{21} \nabla \psi \right) + \nabla \cdot \left(\frac{k_0}{\mu} \nabla p \right). \nonumber
\end{align}
These coupled equations can be written as a matrix differential equation. In the derivation we assume $\sigma_0$, $k_0$, $\mu$, $L_{12}$, and $L_{21}$ are piecewise constant in space (allowing them to be taken outside the divergence operator), which results in the form
\begin{equation}
  \label{eq:pdes-vector-matrix-GP}
  \begin{bmatrix}C^{\ast} & 0 \\ 0 & nc\end{bmatrix} \frac{\partial}{\partial t} 
  \begin{bmatrix} \psi \\ p \end{bmatrix} = 
  \begin{bmatrix} \sigma_0  & L_{12}  \\
                         L_{21} & \frac{k_0}{\mu} \end{bmatrix} \nabla^2  
   \begin{bmatrix} \psi \\ p \end{bmatrix}.
\end{equation}
The approach requires assumption of piecewise constancy in material parameters which can violate flux continuity at property discontinuities in numerical models with linear interpolation functions, but still allows either layers or regions of different material properties. The limitations of the approach are summarized in Section~\ref{sec:limits}.

This can be additionally re-arranged, pre-multiplying by the diagonal storage properties matrix inverse, leaving
\begin{equation}
  \label{eq:pdes-vector-matrix3}
  \frac{\partial \bm{d}}{\partial t} = 
  \begin{bmatrix} \alpha_{E}  & \alpha_{E} K_{S} \\
                 \alpha_H K_E & \alpha_{H} \end{bmatrix} \nabla^2  \bm{d},
\end{equation}
where $\bm{d}=[\psi, p]^T$ is the physical potential vector, $\alpha_{H}=k_0/(\mu n c)$ is the hydraulic diffusivity $\mathrm{[m^2/sec]}$, $\alpha_{E}=\sigma_0/C^{\ast}$ is the electrostatic potential diffusivity $\mathrm{[m^2/sec]}$, $K_S=L_{12} /\sigma_0$ is the streaming potential  $\mathrm{[V/Pa]}$, and $K_E=L_{21}\mu/k_0$ is the electroosmosis pressure $\mathrm{[Pa/V]}$.

Based on a thermodynamic argument of microscopic reversibility, \citet{onsager31,onsager31b} showed the off-diagonal coupling coefficients in expressions like \eqref{eq:pdes-vector-matrix-GP} are symmetric $(L_{12}=L_{21})$ when the fluxes and forces are written correctly  \citep{luikov75,corapcioglu91}. With manipulation, the dimensions of $L_{12}$ $\mathrm{[A/Pa \cdot m]}$ and $L_{21}$ $\mathrm{[m^2/V \cdot sec]}$ can be shown to be equivalent, indicating the equations are written in the correct form. Using this equality, we express permeability in terms of the streaming potential and the electroosmotic pressure $K_E=K_S\sigma_0\mu/k_0$ or $k_0=\sigma_0 \mu\frac{K_S}{K_E}$ \citep{pengra99,pengra99b}.  When $L_{12}=L_{21}=0$ the streaming potential and electroosmotic pressure cease to exist; zeroing these coefficients makes the matrix in \eqref{eq:pdes-vector-matrix3} diagonal, where the flow and electrostatic problems are independent and uncoupled.

Starting with \eqref{eq:pdes-vector-matrix3}, we multiply the $\psi$ equation by $L_c^2/(\alpha_H \Psi_c)$ and multiply the $p$ equation by $L_c^2/(\alpha_H P_c)$.
Characteristic electrostatic potential ($\Psi_c=P_c K_S$), pressure ($P_c$), time ($T_c=L_c^2/\alpha_H$), and length ($L_c$) are used to re-write the equations in non-dimensional form in terms of $x_D=x/L_c$, $t_D=t/T_c$, $p_D=p/P_c$, $\psi_D=\psi/\Psi_c$, and $\nabla_D^2$ (the dimensionless Laplacian). The governing equation becomes
\begin{equation}
  \label{eq:uncouple2}
  \frac{\partial \bm{d}_D}{\partial t} = \bm{A} \nabla_D^2 \bm{d}_D
\end{equation}
where $\bm{A}=  \begin{bmatrix} \alpha_D  & \alpha_D \\ K_D & 1 \end{bmatrix}$ is the dimensionless matrix of Laplacian operator coefficients, $\alpha_D=\alpha_E/\alpha_H$ is the dimensionless electrical/hydrological diffusivity ratio,  $\bm{d}_D=\left[\psi/\Psi_c, p/P_c \right]^T$ is the dimensionless potential vector, and $K_D=K_E K_S$ is the dimensionless product of the electroosmotic pressure and streaming potential, representing the magnitude of electrokinetic coupling. In the formulation presented here these two dimensionless quantities completely characterize the electrokinetic problem, reducing the problem from four free parameters ($\alpha_H$, $\alpha_E$, $K_S$, $K_E$) to two. This reduction in free parameters does not limit the range of validity of the solution, it properly simplifies the previously over-constrained solution space (two equations in terms of two potentials were related with four parameters). $L_c$ and $P_c$ are chosen from the physical problem configuration (e.g., domain size and applied boundary or initial conditions), while $T_c$ and $\Psi_c$ are specified as part of re-writing the governing equations in dimensionless form (two examples of this approach are presented in Section~\ref{sec:applications}). We follow a variable convention where bold lower-case are vectors and bold upper-case are matrices (see Tables~\ref{tab:notation} and \ref{tab:dimensionless} for notation).

\section{Uncoupling}\label{sec:uncoupling}
The uncoupling approach relies on decomposing the matrix characterizing the operator in the governing equation into a diagonal matrix using its eigenvalues and eigenvectors (i.e., spectral decomposition). While the approach has been used in geophysics to decouple poroelastic wave equations \citep{lo09}, and in quantum mechanics to uncouple second-order differential equations regarding elastic scattering \citep{stechel78,light79}, it has not previously been applied to uncoupling electrokinetic processes.

\subsection{Governing Equations}
The method requires $\bm{A}$ is diagonalizable (i.e., it has a complete, linearly independent set of eigenvectors), so it can be decomposed into its eigenvectors and eigenvalues as  $\bm{A}=\bm{S} {\bm \Lambda}\bm{S}^{-1}$ (e.g., \citet[\S 5.2]{strang88}). A non-repeating set of eigenvalues is also indicative of diagonalizability, but some cases with repeating eigenvalues may still be diagonalizable. $\bm{S}$ is a matrix with the eigenvectors of $\bm{A}$ as columns, and $\bm \Lambda$ is a diagonal matrix with the eigenvalues of $\bm{A}$ along the diagonal, with the eigenvectors and eigenvalues in corresponding order. These are substituted into (\ref{eq:uncouple2}) and the expression is pre-multiplied by $\bm{S}^{-1}$ (assuming $\bm{S}$ and $\bm{S}^{-1}$ are piecewise constant in space and time and can be commuted with derivatives)
\begin{equation}
  \label{matrix-form-all-terms}
  \frac{\partial}{\partial t_D}\bm{S}^{-1}\bm{d}_D=(\bm{S}^{-1}\bm{S}) {\bm \Lambda} \nabla_D^2 \bm{S}^{-1}\bm{d}_D ,
\end{equation}
resulting in $\frac{\partial {\bm \delta}}{\partial t_D} =  {\bm \Lambda} \nabla_D^2 {\bm \delta}$, where ${\bm \delta}={\bm S}^{-1} \bm{d}_D$. This expression can be simplified because $\bm{S} {\bm S}^{-1} = \bm{I}$ is the identity matrix. The two equations are now uncoupled in terms of the newly defined intermediate potential ${\bm \delta}$ because $\bm \Lambda$ is a diagonal matrix.  

For the $2 \times 2$ dimensionless electrokinetic problem, the eigenvectors and eigenvalues are computed from the characteristic equation to be
\begin{align}
  \label{eq:SandLfull}
  \bm{S}&=\begin{bmatrix}
    \frac{2 \alpha_D }{1-\alpha_D-\Delta}&
    -\frac{1}{2K_D}\left(1-\alpha_D-\Delta\right)\\
  1&1\end{bmatrix} \nonumber \\
  {\bm \Lambda} &=
  \begin{bmatrix}
      \frac{2 \alpha_D \left( 1 - K_D\right)}{1+\alpha_D+\Delta}&0\\
    0&\frac{1}{2}\left(1+\alpha_D+\Delta\right)\end{bmatrix}\\
  \bm{S}^{-1}&=\frac{1}{\Delta} \begin{bmatrix}
    -K_D & -\frac{1}{2}\left(1-\alpha_D-\Delta\right) \\
    K_D & \frac{-2 K_D \alpha_D}{1-\alpha_D-\Delta}
    \end{bmatrix},\nonumber
\end{align}
where
\begin{equation}
  \label{eq:Delta}
  \Delta = \sqrt{1+ \alpha_D \left( 4K_D -2 + \alpha_D\right)}.
\end{equation}
The uncoupled solution is computed for the two components of ${\bm \delta}$ (i.e., the intermediate potentials) then the physical solution ($\bm{d}_D$) is found from ${\bm \delta}$ via a matrix-vector multiply.

If the two differential equations for hydraulic and electrical flow were not coupled ($K_E=K_S=0$), $\bm{A}$ would already be a diagonal matrix. Writing the governing equations in the one-way coupled form (i.e., including effects of streaming potential, but not including electroosmosis -- or vice-versa) results in a degenerate system of equations which cannot be decoupled using the eigenvalue approach. This one-way coupling approach, although common \citep{casagrande49,banerjee80,malama09-unconfinedSP,malama09-confinedSP} and approximately correct, is physically inconsistent. Since thermodynamics requires $L_{12}=L_{21}$, setting only one of these coefficients to zero breaks the symmetry identified by \citet{onsager31,onsager31b}.

Benefitting numerical evaluation, $\Delta$ \eqref{eq:Delta} will never be complex for positive $\alpha_D$ and $K_D$. For a typical aquifer $K_D \ll 1$ (i.e., $K_S \approx 10^{-8} \; \mathrm{[V/Pa]}$ and $K_E \approx 10^2 \; \mathrm{[Pa/V]}$; \citep{pengra99}), the eigenvalues ($\Lambda_{11}$ and $\Lambda_{22}$) will both be positive. The eigenvalues correspond to the permeability or hydraulic conductivity coefficients used in existing numerical hydrologic simulators. When the eigenvalues are positive existing numerical simulators (e.g., MODFLOW \citep{harbaugh05} or PFLOTRAN \citep{Hammond14}) can be used to solve for the intermediate potentials. The eigenvalues would be substituted for the permeability or hydraulic conductivity, while the storage coefficient or compressibility would be set to unity (depending on the details and units expected in the implementation). Physics precludes negative eigenvalues.

The forms of the uncoupled coefficients \eqref{eq:SandLfull} are written to reduce catastrophic numerical cancellation due to subtraction of like-sized terms. If the naive quadratic formula is used (e.g., \texttt{Eigensystem[]} returned by Mathematica \citep{Mathematica}), severe cancellation will occur because $\alpha_D$ and $\Delta$ may differ by one part in $10^{10}$, rendering this difference numerically inaccurate. For example, the first eigenvalue could be equivalently written $\Lambda_{11}=\frac{1}{2} \left( 1+\alpha_D -\Delta \right)$, but this would need to be computed in quadruple-precision to get a double-precision accurate result (since $\alpha_D \approx \Delta$).

To compute $p_D$ and $\psi_D$ from the intermediate  ${\bm \delta}$ components, pre-multiply by $\bm{S}$,  $\bm{d}_D=\bm{S}{\bm \delta}=\bm{S}\bm{S}^{-1}\bm{d}_D$. The expression for ${\bm \delta}$ in terms of physical variables is used to express source terms or boundary conditions from the intermediate problem in terms of the source terms or boundary conditions of the physical problem, namely
\begin{align}
  \label{eq:delta_written_out}
  {\bm \delta}=\begin{bmatrix} \delta_1 \\ \delta_2 \end{bmatrix}&=\frac{1}{\Delta}\begin{bmatrix}
    -H p_D - K_D\psi_D\\
     G p_D + K_D\psi_D\\
 \end{bmatrix}, 
\end{align}
where
\begin{align}
  \label{eq:Gs}
  G &=\frac{-2 K_D \alpha_D}{1-\alpha_D-\Delta} &
  H &=\frac{1}{2}\left( 1-\alpha_D-\Delta\right).
\end{align}
The expression for $G$ has also been re-written to minimize cancellation between $\alpha_D$ and $\Delta$. The physical potentials ($\bm{d}_D$) are given in terms of intermediate variables as
\begin{align}
  \label{eq:dD_written_out}
  \bm{d}_D=\begin{bmatrix}\psi_D \\ p_D \end{bmatrix}&=\begin{bmatrix}
    \frac{-1}{K_D} \left(G \delta_1 + H \delta_2\right) \\
    \delta_1 + \delta_2\\
 \end{bmatrix}, 
\end{align}
which is used to compute the final physical result from the intermediate representation.

The solution procedure begins by first computing boundary conditions for the intermediate problem from the physical problem, then solving two intermediate uncoupled diffusion problems, and finally recombining the intermediate results to obtain the physical potentials.  The physical properties (i.e., diffusivities) for the two intermediate problems are given by $\Lambda_{11}$ and $\Lambda_{22}$ (eigenvalue matrix diagonals). The boundary conditions or source terms for the intermediate potentials (i.e., in terms of $\bm \delta$) are determined from \eqref{eq:delta_written_out}. Finally, the physical potentials are found from the intermediate potentials by a matrix-vector multiplication \eqref{eq:dD_written_out}.

\subsection{Boundary Conditions}
To solve the intermediate differential equations, any inhomogeneous boundary and initial conditions must also be expressed in terms of ${\bm \delta}$.  The most general inhomogeneous Robin (Type III) boundary condition that can be accommodated by the uncoupling approach is
\begin{align}
  \label{eq:Robin-d}
  \hat{n} \cdot d_c \nabla_D \bm{d}_D = n_c  \bm{d}_D + \bm{c},
\end{align}
where $\hat{n}$ is the boundary normal unit vector, $d_c$ is the dimensionless scalar Dirichlet coefficient, $n_c$ is the dimensionless scalar Neumann coefficient, and $\bm{c}$ is a dimensionless inhomogeneous constant vector.  By setting either $d_c$ or $n_c$ to zero, an inhomogeneous Dirichlet (Type~I) or Neumann (Type~II) condition can be specified. By setting $\bm c$ to zero, a homogeneous boundary condition can be specified. One equation can be homogeneous, and the other inhomogeneous, by setting only one term in $\bm c$ to zero. Pre-multiplying through by $\bm{S}^{-1}$ gives the boundary condition for the intermediate potentials in terms of ${\bm \delta}$ as
\begin{align}
  \label{eq:Robin-delta}
  \hat{n} \cdot d_c \nabla_D {\bm \delta} =  n_c {\bm \delta} + {\bm \gamma},
\end{align}
where ${\bm \gamma}=\bm{S}^{-1}{\bm c}$ is the transformed inhomogeneous term, of the same form as \eqref{eq:delta_written_out}.

\subsection{Extension}
This same uncoupling approach could be applied to larger systems of equation, using Cramer's rule to algebraically solve the resulting eigenvalue problem \citep{strang88}. Coupled processes may include temperature, electrostatic potential, and solute concentration \citep{luikov75,corapcioglu91,leinov14}. The $2\times2$ formulation presented here could readily be extended to a $3\times3$ system including thermal flux and temperature gradient driving forces. Minimizing cancellation in the expressions could become much more tedious with larger systems, but automatic term re-writing software (e.g., Herbie \citep{panchekha15}) could automate this. For larger coupled equation systems, the same approach should still be possible, but the uncoupling could more efficiently be done numerically (e.g., using eigenvalue and eigenvector routines in LAPACK \citep{lapack90}). 

\section{Applications}\label{sec:applications}
We present two applications to demonstrate the eigenvalue uncoupling approach for electrokinetics. The first is one-dimensional cylindrically symmetric flow to a well in an infinite planar aquifer and the second is Cartesian one-dimensional flow driven by harmonic source terms on one end of a finite domain.

\subsection{Streaming Potential from Pumping a Well}\label{section:theis}
The application geometry is a fully penetrating line-source well in a confined aquifer with electrically insulating aquicludes (i.e., no hydraulic or electric flow) above and below. This application is not chosen for its physical realism to field streaming potential applications, but as a simple example using a well-known solution. The characteristic length is the aquifer thickness ($L_c=b$), the characteristic pressure is derived from the specified volumetric pumping rate ($P_c=\mu Q_T/(4 \pi b k_0)$).

Appendix~\ref{sec:Theis-appendix} presents the original and dimensionless form of the governing equations, boundary conditions, and initial conditions.  The intermediate governing equations and initial/boundary conditions (\ref{eq:theis-PDE-UC} \& \ref{eq:theis-BC-UC}) can be solved with the \citet{theis35} solution, defined as the exponential integral of the first kind, a well-known solution for flow to a well in a confined 2D aquifer of infinite radial extent ($T\nabla^2s=S\frac{\partial s}{\partial t}$) \citep{batu98,lee99},
\begin{align}
  \label{eq:theis-soln}
  s = \frac{Q_T}{4 \pi T} \mathrm{E}_1\left(\frac{r^2 S}{ 4 T t }\right),
\end{align}
where $s$ is drawdown (i.e., change in hydraulic head) $\mathrm{[m]}$, $T$ is transmissivity $\mathrm{[m^2/sec]}$, and $S$ is the dimensionless storage coefficient (commonly referred to as storativity).

Substituting the relevant parameter definitions ($s \rightarrow \delta_i$, $T \rightarrow \Lambda_{ii}$, $S \rightarrow 1$, $Q_T \rightarrow Q_i$), the solution to the intermediate potentials can be expressed simply in the form used in ``type-curve'' analysis on log-log scale plots (e.g., \citet{batu98}), 
\begin{align}
  \label{eq:type}
  \mathrm{x\;axis:} && \overbrace{\ln\left[ \frac{t}{r^2} \right]}^{\mathrm{data}} &= \overbrace{\ln\left[\frac{1}{4 \Lambda_{ii}} \right]}^{\mathrm{shift}} + \overbrace{\ln\left[ \frac{1}{u}\right]}^{\mathrm{type-curve}}  \nonumber \\
  \mathrm{y\;axis:}  && \ln\left[ \delta_i \right] &= \ln\left[ \frac{Q_i}{4 \pi \Lambda_{ii}}\right] + \ln\left[ \mathrm{E}_1(u)\right] 
\end{align}
where the first right-hand side term of \eqref{eq:type} represents the $x$ or $y$ type curve shift, $Q_i$ is computed from the physical source terms using \eqref{eq:delta_written_out} and \eqref{eq:Robin-delta}, and $u=r^2S/\left(4Tt\right)=r^2/\left(4 \Lambda_{ii} t \right)$ is the dimensionless argument to $\mathrm{E_1}$. For the uncoupled electrokinetic problem, both axes are shifted in the log-log plot by the intermediate diffusivity ($1/4 \Lambda_{ii}$), while the $y$-axis is shifted additionally by the intermediate source term ($Q_i/\pi$).

\begin{figure}
    \includegraphics[width=0.49\textwidth]{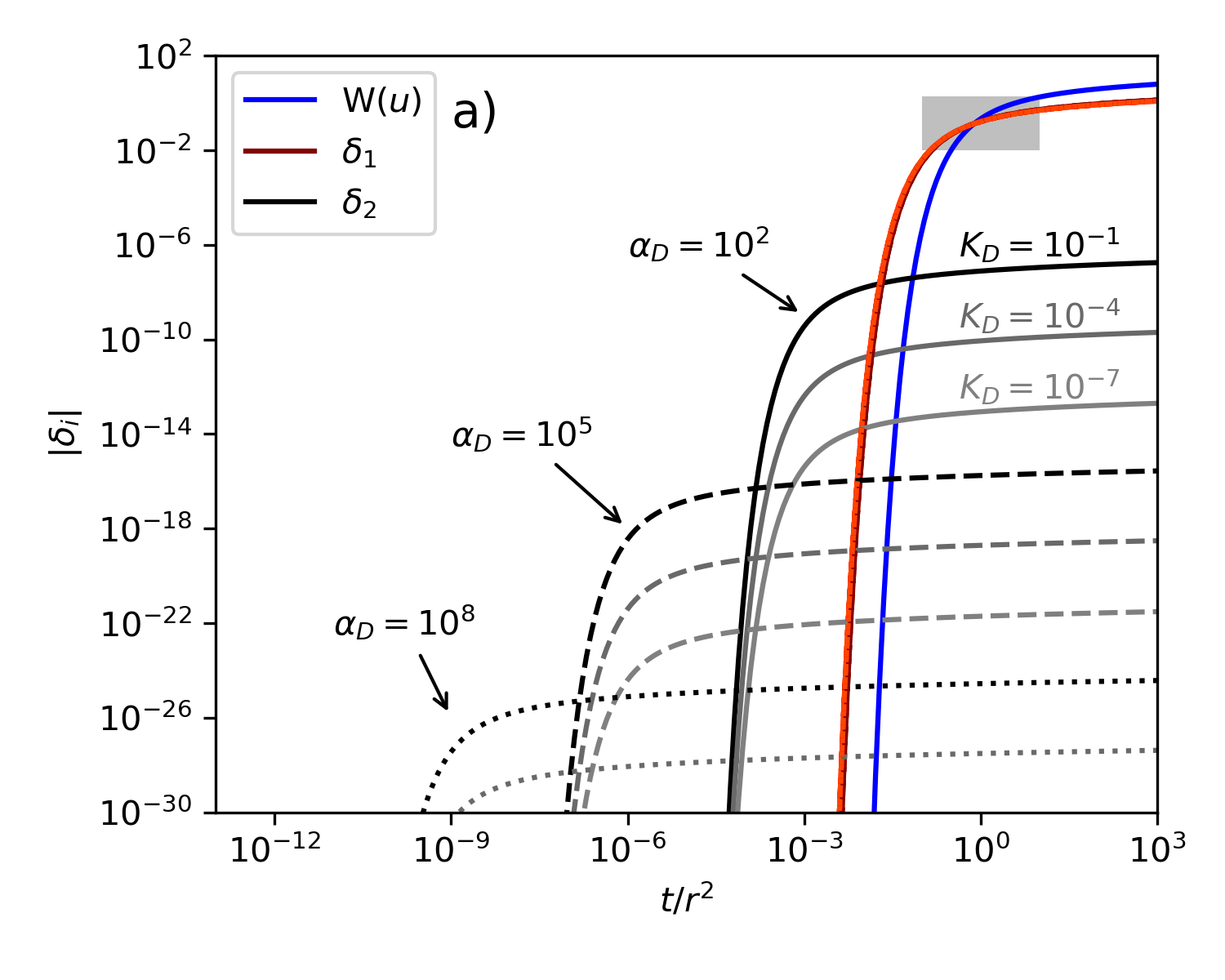}
    \includegraphics[width=0.49\textwidth]{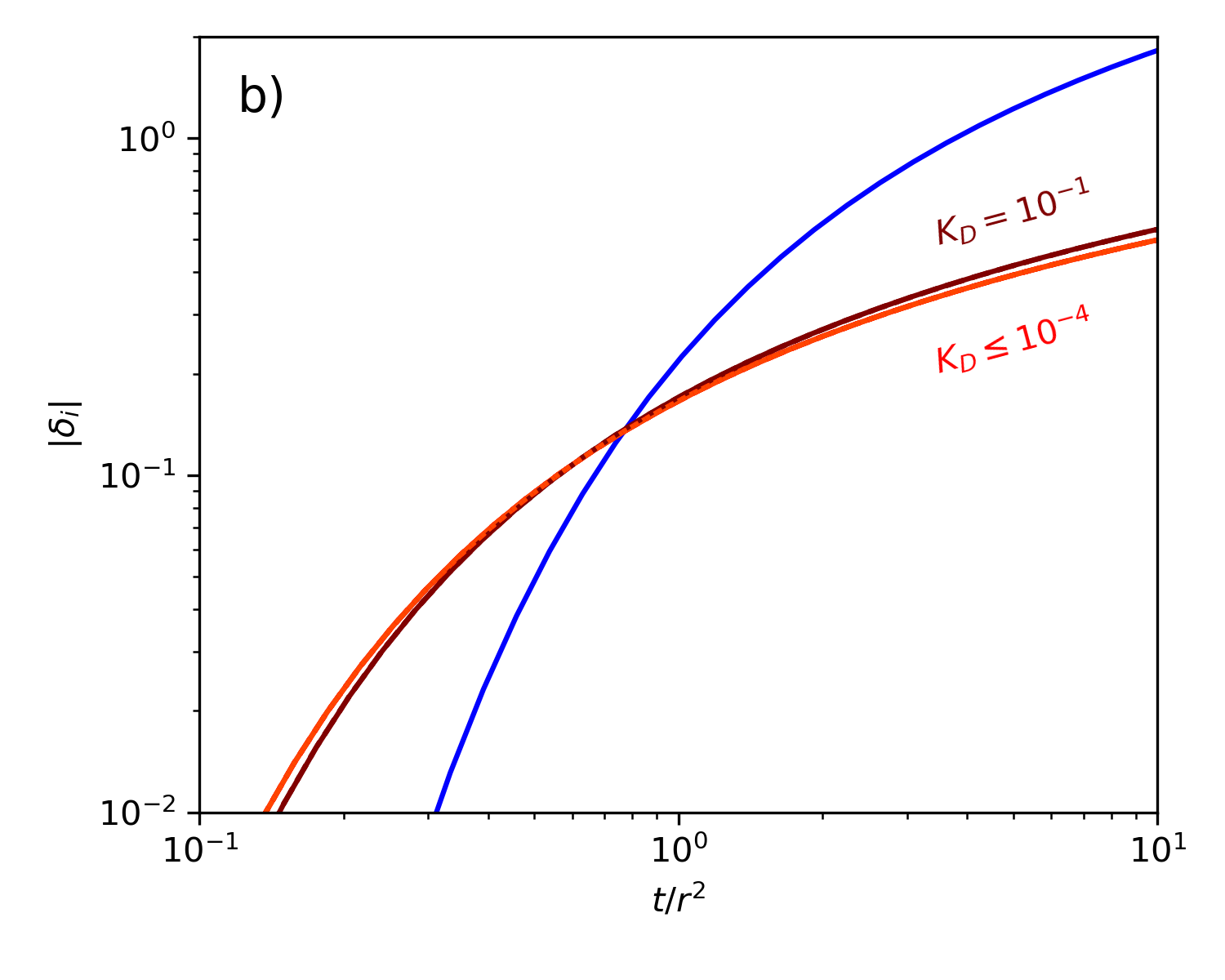}
    \caption{a) shifted Theis solutions for $\delta_i$ for $\alpha_D=\left[ 10^2, 10^5, 10^8\right]$ (corresponding to solid, dashed, and dotted lines, respectively) and $K_D=\left[ 10^{-1}, 10^{-4}, 10^{-7}\right]$ (for dark, medium, and light colored lines, respectively). Shades of red are $\delta_1$ (darkest red is largest $K_D$ value), shades of black are $\delta_2$ (black is largest $K_D$), and the single blue curve is un-shifted $\mathrm{E_1}$ type curve. Subplot b) is zoomed in to a small region in a) indicated with box in a). Different colored $\delta_1$ curves in a) are nearly coincident.}
  \label{fig:theis_deltas}
\end{figure}

Figure~\ref{fig:theis_deltas} shows type-curve $\delta_i$ solutions for several $\alpha_D$ and $K_D$ values. Figure~\ref{fig:theis_deltas}b shows a zoomed-in region, where the small amount of separation between the curves for different values of $K_D$ is visible. The solution was computed with the variable-precision library \texttt{mpmath} \citep{mpmath} to confirm the dependence of the solution on numerical precision, but when terms are written correctly the solution can be computed with double precision.

The parameter range chosen has a much larger effect on $\delta_2$ than on $\delta_1$. Table~\ref{tab:theis-params} shows $\Delta \approx \alpha_D$ and $\Lambda_{11} \approx 1$ for the $\alpha_D$ and $K_D$ range chosen, while $\Lambda_{22} \approx \alpha_D$. Increasing $\alpha_D$ shifts the $\delta_i$ type curves both down and to the left, while decreasing $K_D$ only shifts the curves down. The impact of changing $K_D$ on the $Q_2$ value (only included in the $y$-shift for $\delta_2$) is evident in Table~\ref{tab:theis-params}, while $Q_1$ values do not change appreciably with $K_D$, because $H$ is much larger in magnitude than $G$ in \eqref{eq:delta_written_out} ($G \approx K_D$ and $H \approx -\alpha_D$). Table~\ref{tab:theis-params} shows source terms are $Q_1 \approx 2$ and $Q_2 \approx K_D/\alpha_D^2$.  The physical solution for $p_D$ and $\psi_D$ are obtained by substituting the intermediate solutions ($\delta_i$) into \eqref{eq:dD_written_out}. 

\begin{table}[]
  \caption{Parameters specified (left) and computed (right) for Theis example (six non-zero significant figures).}
  \label{tab:theis-params}
  \begin{tabular}{cc|ccccccc}
    \hline
    $\alpha_D$ & $K_D$     & $\Delta$  & $\Lambda_{11}$ & $\Lambda_{22}$ & $G$                  & $H$     & $Q_1$    &  $Q_2$ \\
    \hline
    $10^{2}$   & $10^{-1}$ & $99.2018$ &$0.899093$      &$100.101$       &$1.00907\times10^{-1}$ &$-99.1009$ &$1.99998$ &$1.8291\times10^{-5}$ \\
    $10^{2}$   & $10^{-4}$ & $99.0002$ &$0.999899$      &$100$           &$1.0101\times10^{-4}$  &$-99.0001$ &$2$       &$2.0404\times10^{-8}$ \\
    $10^{2}$   & $10^{-7}$ & $99$      &$1$             &$100$           &$1.0101\times10^{-7}$  &$-99$      &$2$       &$2.04061\times10^{-11}$ \\
    \hline
    $10^{5}$   & $10^{-1}$ & $99999.2$ &$0.899999$      &$10^5$          &$1.00001\times10^{-1}$ &$-99999.1$ &$2$       &$1.80003\times10^{-11}$ \\
    $10^{5}$   & $10^{-4}$ & $99999$   &$0.9999$        &$10^5$          &$1.00001\times10^{-4}$ &$-99999$   &$2$       &$1.99984\times10^{-14}$ \\
    $10^{5}$   & $10^{-7}$ & $99999$   &$1$             &$10^5$          &$1.00001\times10^{-7}$ &$-99999$   &$2$       &$2.00004\times10^{-17}$ \\
    \hline
    $10^{8}$   & $10^{-1}$ & $10^8$    &$0.9$           &$10^8$          &$10^{-1}$              &$-10^8$    &$2$       &$1.8\times10^{-17}$ \\
    $10^{8}$   & $10^{-4}$ & $10^8$    &$0.9999$        &$10^8$          &$10^{-4}$              &$-10^8$    &$2$       &$1.9998\times10^{-20}$ \\
    $10^{8}$   & $10^{-7}$ & $10^8$    &$1$             &$10^8$          &$10^{-7}$              &$-10^8$    &$2$       &$2\times10^{-23}$ \\
    \hline
  \end{tabular}
\end{table}

\begin{figure}
  \includegraphics[width=\textwidth]{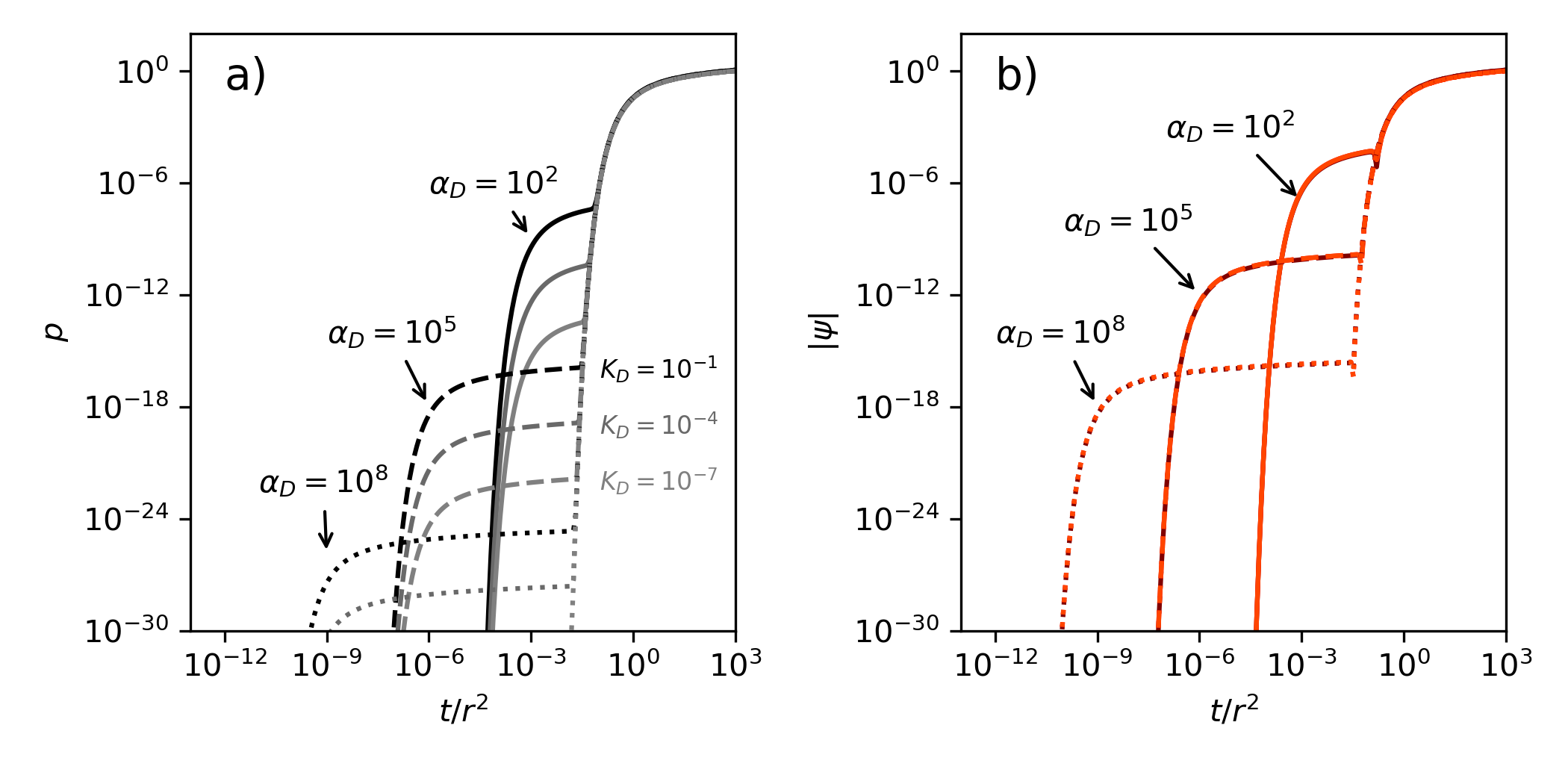}
  \caption{Recombined physical domain solutions (a: pressure response, b: magnitude of electrical response), computed from $\delta_i$  for $\alpha_D=\left[ 10^2, 10^5, 10^8\right]$ (solid, dashed, and dotted lines, respectively) and $K_D=\left[ 10^{-1}, 10^{-4}, 10^{-7}\right]$ (for dark, medium, and light colored lines, respectively) -- different colored curves representing different values of $K_D$} in b) are nearly coincident.
  \label{fig:theis_recombined}
\end{figure}

At early time in Figure~\ref{fig:theis_recombined} the electrical response (b) is larger in magnitude than the pressure response (a). The electrical response at early time ($t/r^2 < 0.01$) is surpassed at later time by the streaming potential response ($\psi$ and $p$ about the same magnitude). The hydraulic diffusivity controls the propagation speed in the streaming potential response; the early response due to electrical conduction from the well is essentially instantaneous for $\alpha_D \gg 100$. The rate of change of the time derivative in the electrical problem has reached steady-state (i.e., the curves are largely horizontal on log-log plots) before the streaming potential response surpasses the electrical conduction from the source well.

The balance between the time scales associated with the electrical ($L^2_c/\alpha_E=b^2 C^\ast/\sigma_0$) and flow ($L^2_c/\alpha_H=b^2 n c \mu /k_0$) problems is captured in $\alpha_D$ (solid vs. dashed vs. dotted lines). The characteristic time scale for the electrical problem is much shorter than the hydrologic problem for $\alpha_D \gg 100$, which is typical. The balance between the primary (diagonal) and coupled (off-diagonal) processes is captured in $K_D$ (darker vs. lighter colors). Larger $K_D$ results in larger electroosmotic pressure due to an applied electrical source.

\begin{figure}
  \includegraphics[width=0.49\textwidth]{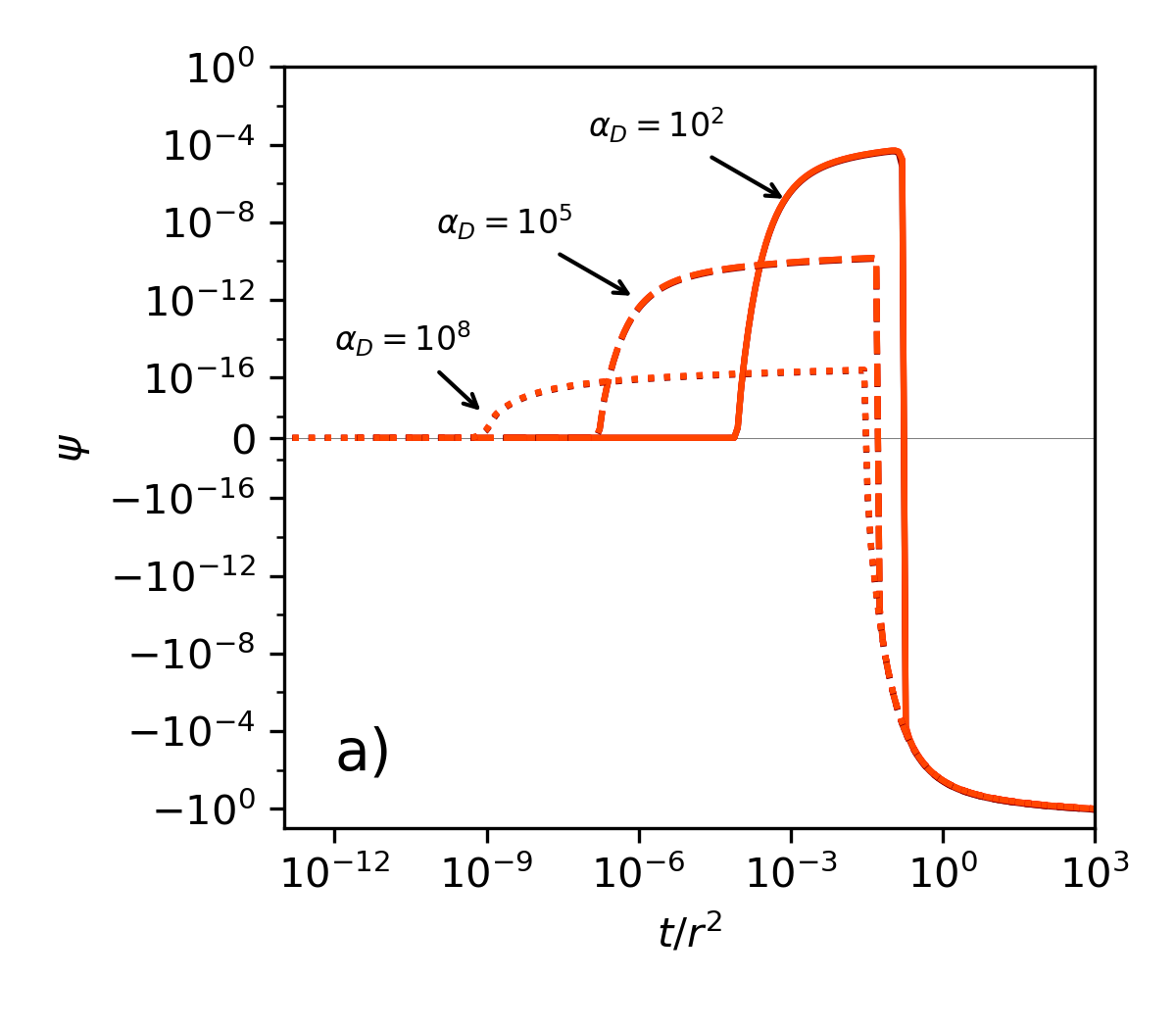}
  \includegraphics[width=0.49\textwidth]{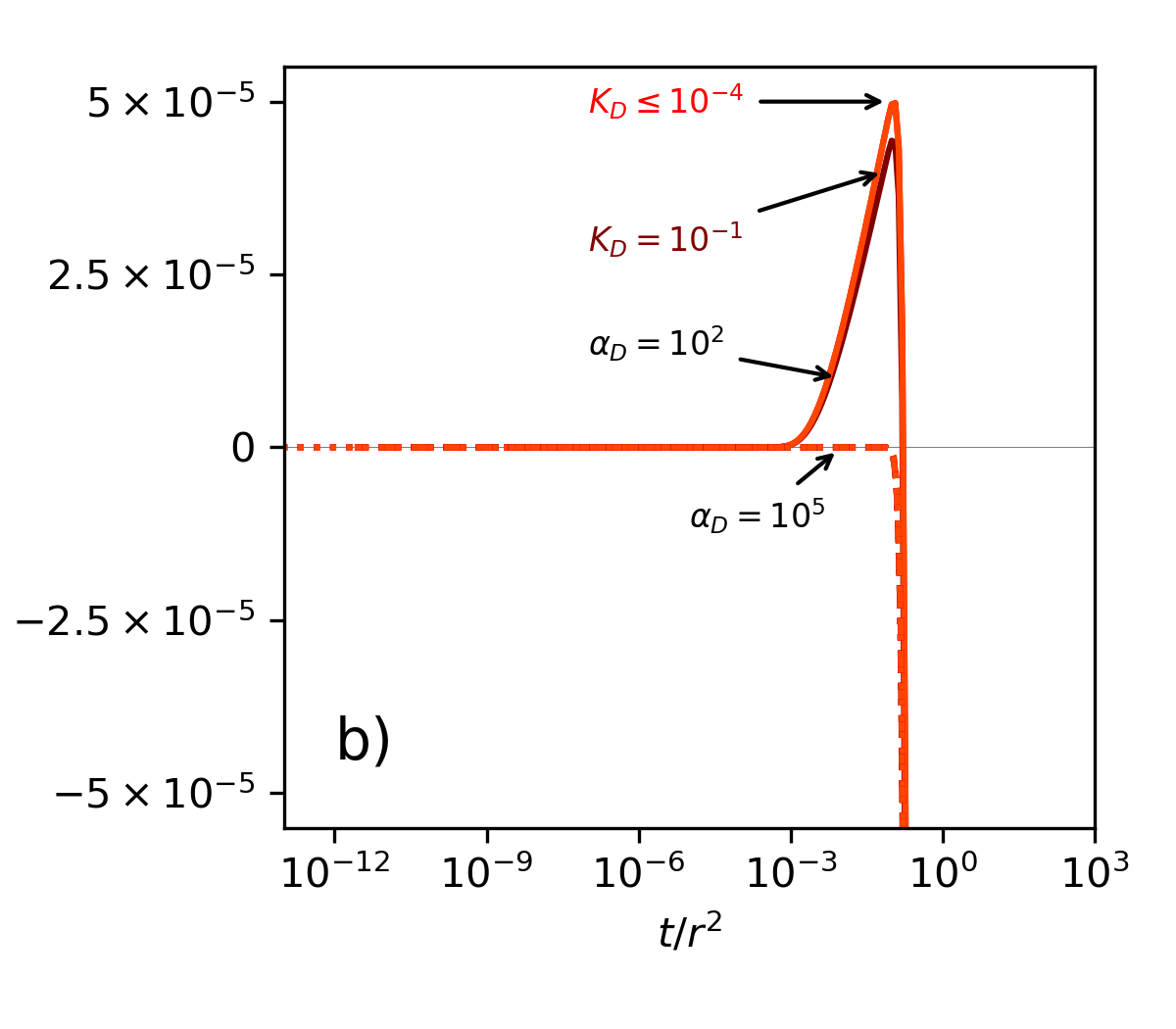}
  \caption{Recombined physical domain electrical solution on a) linear-log $y$-axis scale (linear $-10^{-18} \le y \le 10^{-18}$ and log elsewhere) and b) linear $y$-axis to illustrate sign reversal of signal around $0.01 \le t/r^2 \le 0.1$. Line type indicates $\alpha_D$ value, $K_D$ value indicated by color (curves for different $K_D$ values are nearly coincident in a). Same curves are plotted as in Figure~\ref{fig:theis_recombined}b.}
  \label{fig:theis_reversal}
\end{figure}

Figure~\ref{fig:theis_recombined}a shows the pressure response ($p$) due to early-time conduction from the electrostatic source term at the well (electroosmosis). This response is sensitive to both $\alpha_D$ and $K_D$ (since it is from coupling), while the electrical response ($\psi$ in Figure~\ref{fig:theis_recombined}b) at early time is not sensitive to $K_D$, as it arises directly from an electrical source, not a coupled source (the streaming potential response occurs later, around $t_D=1$). The electrical response is plotted again in Figure~\ref{fig:theis_reversal} with a combination log-linear-log $y$-axis a) and a linear $y$-axis b) clearly illustrating the sign reversal. For this choice of source magnitudes (i.e., the signs of the terms in \eqref{eq:theis-BC}), negative $\psi$ responses arise from electrical conduction, while larger positive $\psi$ responses comes from streaming potential.


For comparison, the one-dimensional radially symmetric flow problem was solved fully coupled and via eigenvalue decoupling using the finite-volume python multiphysics library \texttt{fipy} \citep{FiPy:2009}. The infinite radial domain was approximated with a large radial domain ($r>2.5$ km) with a non-uniform mesh (550 elements, starting at $\Delta r=0.01$ m, each element growing 2\% over its neighbor) and insulating/no-flow far-field boundary conditions. The dimensionless problem was solved using typical double-precision variables, using the $\alpha_D$ and $K_D$ range shown in Table~\ref{tab:theis-params}. The line source term at the origin is approximated as a 0.02 cm diameter source.
\begin{figure}
  \includegraphics[width=\textwidth]{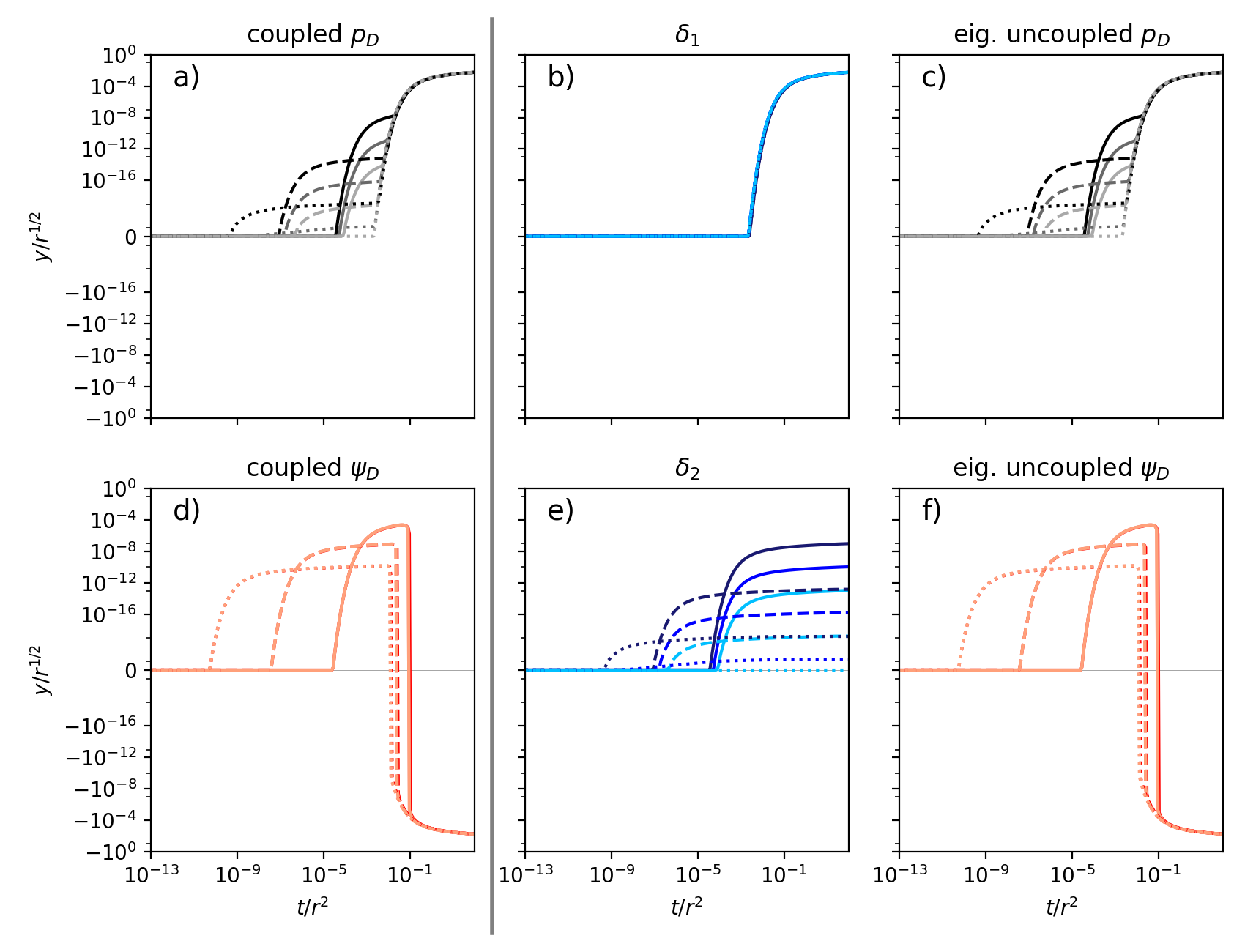}
  \caption{Scaled \texttt{fipy} \citep{FiPy:2009} finite-volume approximation of type-curve example. Subplots a) and d) show fully coupled solution, b) and e) show intermediate ($\delta_i$) solution, c) and f) show re-combined  physical solution. Solutions are for $\alpha_D=\left[ 10^2, 10^5, 10^8\right]$ (for solid, dashed, and dotted lines, respectively) and $K_D=\left[ 10^{-1}, 10^{-4}, 10^{-7}\right]$ (for dark, medium, and light colored lines, respectively).}
  \label{fig:theis_fipy}
\end{figure}

The fully coupled finite-volume problem (Figure~\ref{fig:theis_fipy}a and d) produced similar results to the Theis analytical solution eigenvalue uncoupling approach (Figure~\ref{fig:theis_recombined} but using double-precision limits plotting the solution on the same scale used by the analytical type-curve method). For more direct  comparison, the fipy framework was also used to solve the intermediate problems Figure~\ref{fig:theis_fipy}b and e, and after recombining the two finite-volume numerical solutions were compared directly Figure~\ref{fig:theis_fipy}c and f. Only for $\alpha_D > 10^{12}$ was there any difference between the two approaches, and then the maximum relative difference was $\approx 3\times 10^{-4}$, occurring near the polarity reversal around $t/r^2 \approx 0.01$.

This example illustrates the eigenvalue uncoupling approach, using simplistic type-curve analysis to the predict coupled physics of electrokinetics. The approach could be extended to different aquifer or pumping configurations \citep{maineult08,malama09-confinedSP,malama09-unconfinedSP,malama2014theory} using existing diffusion models as building blocks.

\subsection{Harmonically Driven Laboratory Test}\label{sec:core}
Periodically driven electrokinetics have seen some interesting laboratory applications \citep{pengra99,pengra99b,peng2019effect}, including oscillatory streaming potential \citep{reppert01,tardif11,jouniaux12,glover12a,glover12b,glover2020experimental} and electroosmosis \citep{reppert02}. In oscillatory systems, the coupling coefficient dependence on frequency becomes important \citep{reppert01}. This example demonstrates the eigenvalue uncoupling method for a simple-to-derive analytical solution, as an alternative to more general coupled numerical approaches.

The low-frequency sinusoidal (i.e., no magnetism) driven problem is Cartesian one-dimensional, with a periodic Type-I boundary condition at $x=1$, and a homogeneous Type-II at $x=0$. We solve for the late-time periodic solution using separation of variables.  The governing equations for flow in a one-dimensional domain are derived in Appendix~\ref{sec:oscillatory-appendix}.

Figure~\ref{fig:U} shows the real and imaginary parts of $U$ (the complex wave amplitude). The real portion represents the amplitude while the imaginary portion shows the phase shift between the boundary condition and points inside the domain. The absolute value of $U$ shows the largest magnitude occurs along $x_D=1$ and $\omega_D=0$ (left part of Figure~\ref{fig:U}), corresponding to the location of the harmonically driven boundary condition and at the low-frequency limit. The phase of $U$ (right part of Figure~\ref{fig:U}) shows the response in the domain is in-phase with the boundary conditions at the same locations where it is largest in magnitude ($x_D=1$ and $\omega_D=0$), and at the far side of the domain ($x_D=0$) and at higher frequency ($\omega_D>10$) the solution is significantly out-of-phase with the harmonically driven boundary conditions.  

\begin{figure}
  \includegraphics[width=0.46\textwidth]{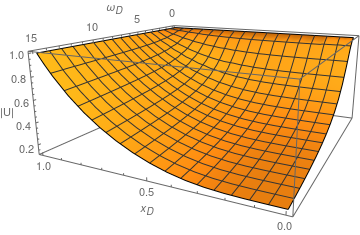}
  \includegraphics[width=0.51\textwidth]{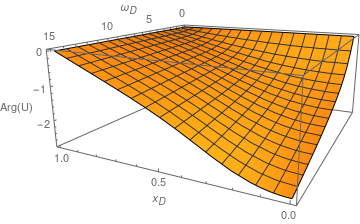}
  \caption{Absolute value (left) and phase (right) of $U(x_D)$, showing frequency ($0 \le \omega_D \le 15$) and space ($0 \le x_D \le 1$) dependence for driven Type-I boundary condition at $x_D=1$ and homogeneous Type-II at $x_D=0$.}
  \label{fig:U}
\end{figure}

\begin{figure}
  \includegraphics[width=0.48\textwidth]{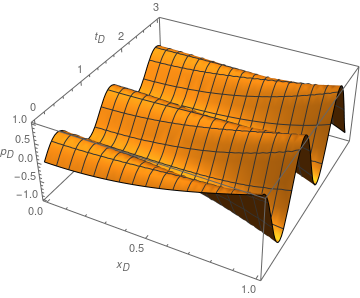}
  \includegraphics[width=0.48\textwidth]{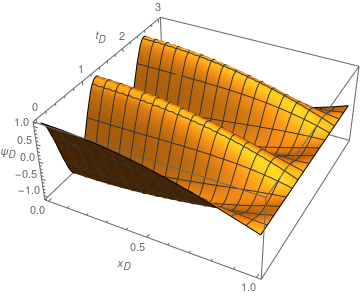}
  \caption{Predicted dimensionless streaming potential pressure ($p_D$ left) and electrical ($\psi_D$ right) response for $\omega_D=5$, $K_D=0.01$, and $\alpha_D=10$. Domain has driven Type-I boundary condition at $x_D=1$, homogeneous Type-II at $x_D=1$.}
  \label{fig:1DSP}
\end{figure}

Figure~\ref{fig:1DSP} shows the dimensionless pressure ($p_D$ left) and electrical ($\psi_D$ right) responses in space and time reconstructed using eigenvalue decoupling for the parameters $\omega_D=5$, $K_D=0.01$, and $\alpha_D=10$. This problem has an applied sinusoidal pressure boundary condition at $x=1$, with the response of the electrical system due to electrokinetic coupling (i.e., there is no inhomogeneous electrical source term). As given by the behavior of $U$ in Figure~\ref{fig:U}, the left plot in Figure~\ref{fig:1DSP} shows the driven $p_D$ problem is largest in magnitude along the boundary at $x_D=1$, but different from $U$ the right plot in Figure~\ref{fig:1DSP} shows the $\psi_D$ response is zero along this same boundary. The recombination step has taken two periodic analytical solutions ($\delta_i$) that are maximum amplitude at the boundary $x_D=1$ and recombined them to conform to the applied boundary conditions, resulting in one periodic solution and one constant solution there.  The magnitude of the oscillations in $p_D$ decreases moving away from the driven boundary condition at $x_D=1$, while the magnitude of the oscillations in $\psi_D$ increase while moving away from the driven boundary condition (where they are fixed to be zero).

\begin{figure}
  \includegraphics[width=0.48\textwidth]{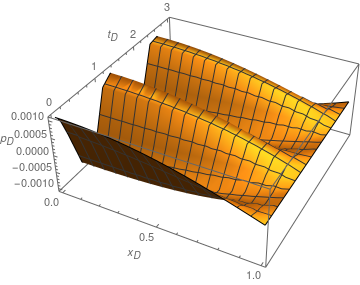}
  \includegraphics[width=0.48\textwidth]{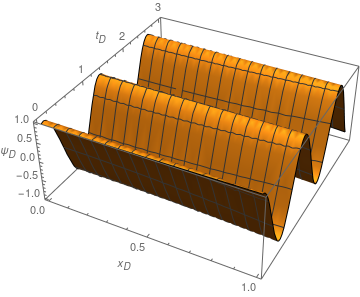}
  \caption{Predicted dimensionless electroosmosis pressure ($p_D$ left) and electrical ($\psi_D$ right) response for $\omega_D=5$, $K_D=0.01$, and $\alpha_D=10$. Domain has driven Type-I boundary condition at $x_D=1$, homogeneous Type-II at $x_D=1$.}
  \label{fig:1DEO}
\end{figure}

Figure~\ref{fig:1DEO} shows the analogous problem for electroosmosis (driven sinusoidal electrical boundary condition at $x_D=1$, passive coupled pressure response). In the electroosmosis problem, the driven $\psi_D$ solution is nearly constant-amplitude across the domain (due to the relatively higher electrical diffusivity), while the $p_D$ solution amplitude increases moving away from the fixed boundary condition at $x_D=1$, but is of relatively small overall magnitude.  

\begin{figure}
  \includegraphics[height=3.5cm]{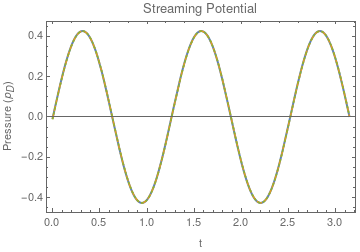}  %
  \includegraphics[height=3.5cm]{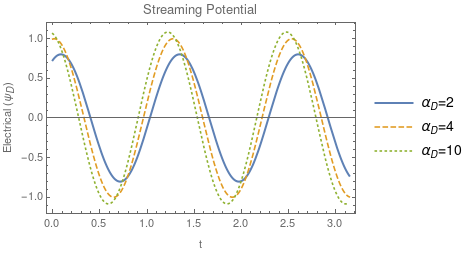}\\
  \includegraphics[height=3.5cm]{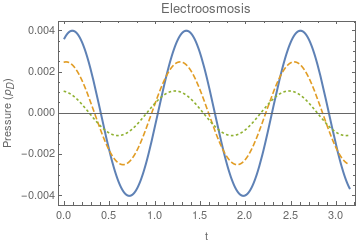}  %
  \includegraphics[height=3.5cm]{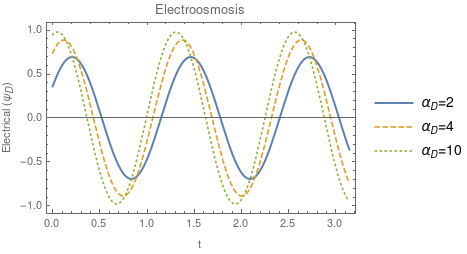}
  \caption{Predicted pressure (left) and electrical (right) responses through time  ($\omega_D=5$,  $K_D=0.01$,  $x_D=0$) for three values of  $\alpha_D$. Domain has specified sinusoidal pressure (streaming potential, top)  or voltage (electroosmosis, bottom) or at  $x_D=1$, no-flow and electrically insulating boundary condition at $x_D=0$.}
  \label{fig:glover}
\end{figure}

\begin{figure}
  \includegraphics[height=5cm]{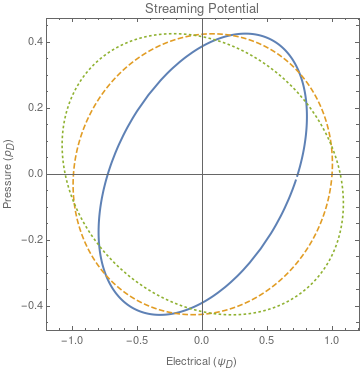}
  \includegraphics[height=5cm]{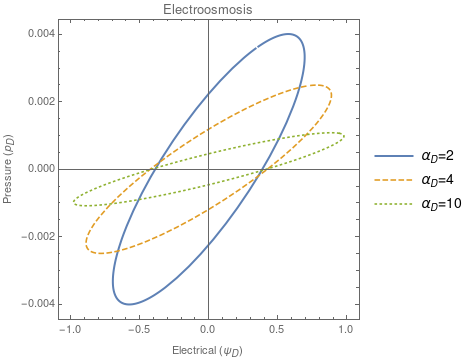}  
  \caption{Parametric plot of predicted electrical ($\psi_D$ abscissa) and pressure ($p_D$ ordinate) response ($\omega_D=5$,  $K_D=0.01$,  $x_D=0$) for three values of  $\alpha_D$ in streaming potential (left) and electroosmosis (right) problems. Same responses as previous figure, but illustrating the sinusoidal parametric form used for fitting data by \citet{glover20EGU}}
  \label{fig:glover-parametric}
\end{figure}

This type of analytical solution (with appropriate physical dimensions and boundary conditions) could be used to fit to laboratory experimental data \protect \citep[e.g.,][]{peng2019effect,glover2020experimental,glover20EGU}. Using the analytical solution illustrated in the previous figures, Figure~\ref{fig:glover} illustrates solutions for three values of $\alpha_D$ for the electroosmosis problem (applied voltage) and the streaming potential problem (applied pressure) through time at $x_D=0$ (left edges of plots in Figures~\ref{fig:1DSP} and \ref{fig:1DEO}).  Figure~\ref{fig:glover-parametric} illustrates the same solutions plotted parametrically against one another for parameter estimation purposes, similar to analyses of laboratory data presented in \citet{glover20EGU}. For the parameters chosen, these figures show the applied pressure problem (streaming potential, top) the measured pressure response does not significantly change for the range of $\alpha_D$, but in the applied voltage problem (electroosmosis, bottom) the measured voltage does change across the range of $\alpha_D$ shown here. For electroosmosis (bottom), the pressure response decrease while the voltage response increases for increasing $\alpha_D$. In the electrical response for both types of tests (right), the response increases with increasing $\alpha_D$. Only in the electroosmosis problem does the pressure response (left) decrease significantly with $\alpha_D$ (different pressure scale in two pressure plots).

\section{Discussion and Method Limitations}\label{sec:limits}
We present an approach for exactly solving certain fully coupled electrokinetic problems, which should also be applicable to other coupled physics problems with symmetric governing equations and boundary conditions. As was shown in the examples, these problems can already be solved using fully coupled numerical models (e.g., Figure~\ref{fig:theis_fipy}) or sometimes approximately using analytical solutions with only one-way coupling \citep{banerjee80,shapiro93,reppert02,malama09-unconfinedSP,malama09-confinedSP,malama2014theory}. The uncoupling approach presented here allows solving the fully coupled problem (i.e., considering both streaming potential and electroosmosis) using simpler numerical or analytical models as computational building blocks. The examples illustrate the ease at which fully coupled electrokinetic analytical solutions can be developed or re-purposed for solving coupled electrokinetic problems.

The limitations of the approach are mostly related to required symmetry. The governing equations must be symmetric (i.e., there must be analogous terms in each equation), which requires including a small transient term in the electrostatic potential equation. The boundary conditions require symmetry between the two physical problems, i.e., the electrical and hydrological equations must have compatible boundary conditions at each spatial location -- they cannot have an specified voltage and specified head gradient conditions at the same location. Homogeneous boundary conditions (i.e., no change in head, no change in voltage, no-flow, or electrically insulated) are simplest, as they are unchanged between the physical and intermediate problems. The decoupling method doesn't require writing the equations in dimensionless form or permeability and electrokinetic coupling coefficients to be constant in space, but if an analytical solution is being used to solve for $\delta_i$, dimensionless problems and homogeneous domains are more straightforward. For heterogeneous domains in numerical models, the approach requires a different set of coupling matricies computed from the flow and electrical properties (${\bm \Lambda}$ and $\bm{S}$) for each region of piecewise-constant material properties, possibly a different set of matrices at each model element. 

Although some care needs to be taken when formulating the coefficients representing material properties and source terms in the intermediate problem to reduce catastrophic cancellation, the formulation of ${\bm \Lambda}$ and $\bm{S}$ presented here is numerically stable and can produce accurate results. While coupled numerical solutions are common and available from commercial finite-element software, analytical solutions are still useful to the hydrologist or geophysicist, since they are often dimensionless, and they do not depend on things of secondary importance to the physical problem, such as mesh resolution, domain size, and time-step size.

\begin{table}[htb]
  \caption{Physical properties}
  \label{tab:notation}
  \begin{tabular}{cll}
    $c$ & compressibility &$\mathrm{[1/Pa]}$ \\
    $C^{\ast}$ & specific capacitance& $\mathrm{[C/(m^3 \cdot V)]}$ \\
    $\bm{j}_e$ & current density vector& $\mathrm{[A/m^2]}$ \\
    $\bm{j}_f$ & Darcy flux vector& $\mathrm{[m/s]}$ \\
    $k_0$ & permeability &$\mathrm{[m^2]}$ \\
    $K_E$ & electroosmosis pressure &$\mathrm{[Pa/V]}$ \\
    $K_S$ & streaming potential &$\mathrm{[V/Pa]}$ \\
    $L_{12}$, $L_{21}$ & electrokinetic coupling coefficients &$\mathrm{[A/(Pa \cdot m)],[m^2/(V \cdot sec)]}$ \\
    $n$ & porosity &$\mathrm{[-]}$ \\
    $p$ & change in fluid pressure &$\mathrm{[Pa]}$ \\
    $t$ & time &$\mathrm{[sec]}$ \\
    $\alpha_E$ & electrical diffusivity &$\mathrm{[m^2/sec]}$ \\
    $\alpha_H$ &  hydraulic diffusivity &$\mathrm{[m^2/sec]}$ \\
    $\mu$ & water viscosity  &$\mathrm{[Pa \cdot sec]}$ \\
    $\psi$ & change in electrostatic potential &$\mathrm [V]$ \\
    $\rho$ & water density & $\mathrm{[kg/m^3]}$ \\
    $\sigma_0$ & electric conductivity & $\mathrm{[S/m]}$ \\
  \end{tabular}
\end{table}

\begin{table}[htb]
  \caption{Dimensionless quantities}
  \label{tab:dimensionless}  
  \begin{tabular}{cll}
    $\bm{A}$ && governing equation coefficient matrix \\
    $\bm{d}_D$ &$\left[\psi_D \; p_D\right]^{T}$& physical potential vector \\
    $K_D$ & $K_EK_S$ & coupling coefficient \\
    $p_D$ & $p/P_c$ & fluid pressure \\
    $r_D$ & $r/L_c$ & radial distance\\
    $\bm{S}$ && eigenvector matrix\\
    $t_D$ & $t/T_c$ & time \\
    $x_D$ & $x/L_c$ & Cartesian distance \\
    $\alpha_{D}$ & $\alpha_E/\alpha_H$ & diffusivity ratio \\
    ${\bm \delta}$ &$\left[\delta_1 \; \delta_2 \right]^{T}$& intermediate potential vector \\
    ${\bm \Lambda}$ && diagonal eigenvalue matrix \\
    $\psi_D$ & $\psi/\Psi_c$ & electrostatic potential \\
  \end{tabular}
\end{table}

\newpage
\section{Declarations}
\subsection{Funding}
This work is funded by the Sandia National Laboratories Earth Science Research Foundation Laboratory-Directed Research and Development (LDRD) program.

\subsection{Conflicts of interest/Competing Interests}
There are no conflicts or competing interests to disclose.

\subsection{Availability of Data and Materials}
The Python and Mathematica scripts used to compute the solutions and make the figures are archived at \texttt{Zenodo.org} with the DOI \texttt{10.5281/zenodo.3979452}. 

\begin{acknowledgements}
This paper describes objective technical results and analysis. Any subjective views or opinions expressed in the paper do not necessarily represent the views of the U.S. Department of Energy or the United States Government.

Sandia National Laboratories is a multimission laboratory managed and operated by National Technology \& Engineering Solutions of Sandia, LLC, a wholly owned subsidiary of Honeywell International Inc., for the U.S. Department of Energy's National Nuclear Security Administration under contract DE-NA0003525.

The authors thank Tara LaForce, Paul Glover, and an anonymous reviewer for their reviews and insightful comments on the manuscript.
\end{acknowledgements}

\bibliographystyle{MG}
{\footnotesize

}

\newpage
\appendix
\section{Cylindrical Flow}
\label{sec:Theis-appendix}
\renewcommand{\theequation}{\thesection.\arabic{equation}}

The governing equations for coupled fluid and electrical response due to pumping an infinitesimal diameter fully penetrating well at constant volumetric flowrate are modified from \citep{malama09-unconfinedSP,malama09-confinedSP}
\begin{align}
  \label{eq:theis-PDE}
  \frac{1}{\alpha_H} \frac{\partial p}{\partial t} &=  K_E \frac{\partial}{\partial r} \left ( r \frac{\partial \psi}{\partial r} \right) +  \frac{\partial}{\partial r} \left ( r \frac{\partial p}{\partial r} \right) , &
  \frac{1}{\alpha_E} \frac{\partial \psi}{\partial t} &= \frac{\partial}{\partial r} \left (  r \frac{\partial \psi}{\partial r} \right) + K_S \frac{\partial}{\partial r} \left (r \frac{\partial p}{\partial r} \right).
\end{align}
The boundary and initial conditions are
\begin{align}
  \label{eq:theis-BC}
  \lim_{r \rightarrow 0} r \frac{\partial p}{\partial r} &= \frac{-Q_T \mu}{2 \pi b k_0} = -2 P_c &
  \lim_{r \rightarrow 0} r \frac{\partial \psi}{\partial r} &= 2 \Psi_c \nonumber \\
  p(r \rightarrow \infty,t) &= 0  &   \psi(r \rightarrow \infty,t) &= 0 \\
  p(r,t=0) &=0 &\psi(r,t=0) & =0,\nonumber
\end{align}
where $Q_T$ is the specified volumetric flowrate $[\mathrm{m^3/sec}]$, $b$ is the aquifer thickness $[\mathrm{m}]$, $r$ is the radial-cylinder coordinate coaxial with the pumping well $[\mathrm{m}]$,  and $t$ is the time since pumping began $[\mathrm{sec}]$. The three constraints in \eqref{eq:theis-BC} specify the constant-flowrate wellbore boundary condition, the far-field no-change condition, and the homogeneous initial condition.

The governing equations, initial conditions, and boundary conditions can be re-written in dimensionless form as
\begin{align}
  \label{eq:theis-PDE-ND}
  \frac{\partial p_D}{\partial t_D} &= K_D \frac{\partial}{\partial r_D} \left ( r_D \frac{\partial p_D}{\partial r_D} \right) + \frac{\partial}{\partial r_D} \left ( r_D \frac{\partial \psi_D}{\partial r_D} \right) \nonumber \\
  \frac{1}{\alpha_D}\frac{\partial \psi_D}{\partial t_D} &= \frac{\partial}{\partial r_D} \left ( r_D \frac{\partial \psi_D}{\partial r_D} \right) + \frac{\partial}{\partial r_D} \left ( r_D \frac{\partial p_D}{\partial r_D} \right)
\end{align}
and
\begin{align}
  \label{eq:theis-BC-ND}
  \lim_{r_D \rightarrow 0} r_D \frac{\partial p_D}{\partial r_D} &= -2 &
  \lim_{r_D \rightarrow 0} r_D \frac{\partial \psi_D}{\partial r_D} &= 2 \nonumber \\
  p_D(r_D \rightarrow \infty,t_D) &= 0  &   \psi_D(r_D \rightarrow \infty,t_D) &= 0 \\
  p_D(r_D,t_D=0) &=0 &\psi_D(r_D,t_D=0) & =0. \nonumber
\end{align}
These coupled dimensionless equations are uncoupled using the approach in Section~\ref{sec:uncoupling}, resulting in the two simpler diffusion equations
\begin{align}
  \label{eq:theis-PDE-UC}
  \frac{\partial \delta_{1}}{\partial t_D} &= \Lambda_{11} \frac{\partial}{\partial r_D} \left( r_D \frac{\partial \delta_{1}}{\partial r_D} \right) & 
  \frac{\partial \delta_{2}}{\partial t_D} &= \Lambda_{22} \frac{\partial}{\partial r_D} \left( r_D \frac{\partial \delta_{2}}{\partial r_D} \right).
\end{align}
The uncoupled boundary and initial conditions are
\begin{align}
  \label{eq:theis-BC-UC}
  \lim_{r_D \rightarrow 0} r_D \frac{\partial \delta_1}{\partial r_D} &= \frac{2}{\Delta} \left(H + K_D \right) \equiv Q_1 &
  \lim_{r_D \rightarrow 0} r_D \frac{\partial \delta_2}{\partial r_D} &= -\frac{2}{\Delta} \left(G + K_D \right) \equiv Q_2 \nonumber \\
  \delta_1(r_D \rightarrow \infty,t_D) &= 0  &   \delta_2(r_D \rightarrow \infty,t_D) &= 0 \\
  \delta_1(r_D,t_D=0) &=0 &\delta_2(r_D,t_D=0) & =0. \nonumber
\end{align}
The wellbore boundary conditions for $\delta_1$ and $\delta_2$ are both inhomogeneous and of different magnitude \eqref{eq:theis-BC-UC}, while for the physical problem the source terms are equal and opposite in sign \eqref{eq:theis-BC-ND}. Homogeneous boundary and initial conditions in the physical domain remain homogeneous in the intermediate domain.

\section{Oscillatory Flow}
\label{sec:oscillatory-appendix}
The coupled equations for electrokinetic flow in a core-scale laboratory apparatus with low-frequency oscillatory boundary conditions are derived, uncoupled, and solved using an analytical solution for a periodic steady-state solution at the same frequency as the applied boundary conditions.

\subsection{Governing Equations}
The governing equations for coupled fluid and electroosmotic response due to applying either sinusoidal voltage (EO) or pressure (SP) at the ends of a one-dimensional column is
\begin{align}
  \label{eq:periodic-PDE}
  \frac{1}{\alpha_H} \frac{\partial p}{\partial t} &=  K_E \frac{\partial^2 \psi}{\partial x^2}  +   \frac{\partial^2 p}{\partial x^2} , &
  \frac{1}{\alpha_E} \frac{\partial \psi}{\partial t} &=  \frac{\partial^2 \psi}{\partial x^2} + K_S  \frac{\partial^2 p}{\partial x^2} 
\end{align}
and the boundary conditions for the applied-pressure SP problem are
\begin{align}
  \label{eq:SP-BC}
  p(x=L) &=F_p \cos\left( \omega t\right)                &    \psi(x=L) &= 0 \nonumber \\
  \left. \frac{\partial p}{\partial x} \right|_{x=0} &= 0 &      \left. \frac{\partial \psi}{\partial x} \right|_{x=0} &= 0
\end{align}
where $F_p$ is the applied pressure signal amplitude $\mathrm{[Pa]}$ and $\omega$ is the applied boundary condition frequency $[\mathrm{1/sec}]$.
The analogous boundary conditions for the applied-voltage EO problem are
\begin{align}
  \label{eq:EO-BC}
  p(x=L) &=0                &    \psi(x=L) &= F_\psi \cos \left( \omega t\right) \nonumber \\
  \left. \frac{\partial p}{\partial x} \right|_{x=0} &= 0 &      \left. \frac{\partial \psi}{\partial x} \right|_{x=0} &= 0
\end{align}
where $F_\psi$ is the applied electrostatic potential signal amplitude $\mathrm{[V]}$.  The domain has an arbitrary initial condition, which is omitted, since it has no influence on the steady-state solution.

Using $L_c=L$, $P_c=F_p$, and $\Psi_c=F_\psi$, the governing equation can be re-written in dimensionless form as
\begin{align}
  \label{eq:periodic-PDE-ND}
  \frac{\partial p_D}{\partial t_D} &=  K_D \frac{\partial^2 \psi_D}{\partial x_D^2}  +   \frac{\partial^2 p_D}{\partial x_D^2} , &
  \frac{1}{\alpha_D}\frac{\partial \psi_D}{\partial t_D} &=  \frac{\partial^2 \psi_D}{\partial x_D^2} + \frac{\partial^2 p_D}{\partial x_D^2}
\end{align}
while the boundary conditions are re-written in dimensionless form as
\begin{align}
  \label{eq:periodic-BC-ND}
  \left. \frac{\partial p_D}{\partial x_D} \right|^{\{\mathrm{SP,EO}\}}_{x_D=0} &=  0 & \left. \frac{\partial \psi_D}{\partial x_D} \right|^{\{\mathrm{SP,EO}\}}_{x_D=0} &= 0 \nonumber \\
  p_D^{\mathrm{SP}}(x_D=1)  &= \cos \left( \omega_D t_D \right) & \psi_D^{\mathrm{SP}}(x_D=1) &= 0 \\
  p_D^{\mathrm{EO}}(x_D=1)  &= 0 & \psi_D^{\mathrm{EO}}(x_D=1) &= \cos \left( \omega_D t_D \right) \nonumber
\end{align}
where $\omega_D=\omega T_c$ is the dimensionless applied frequency for the SP and EO configurations.

These dimensionless equations are uncoupled using the approach in Section~\ref{sec:uncoupling}, resulting in two uncoupled diffusion problems 
\begin{align}
  \label{eq:periodic-PDE-UC}
  \frac{\partial \delta_1}{\partial t_D} &= \Lambda_{11} \frac{\partial^2 \delta_1}{\partial x_D^2} &
  \frac{\partial \delta_2}{\partial t_D} &= \Lambda_{22} \frac{\partial^2 \delta_2}{\partial x_D^2}
\end{align}
while the inhomogeneous boundary conditions are given for the uncoupled intermediate problems (in SP or EO configurations) as
\begin{align}
  \label{eq:periodic-BC-UC}
  \delta^{\mathrm{SP}}_1(x_D=1) &= \xi_1^{\mathrm{SP}} \cos \left( \omega_D t_D \right) & \delta^{\mathrm{SP}}_2(x_D=1) &= \xi_2^{\mathrm{SP}} \cos \left( \omega_D t_D \right), \nonumber \\
    \delta^{\mathrm{EO}}_1(x_D=1) &= \xi_1^{\mathrm{EO}} \cos \left( \omega_D t_D \right) & \delta^{\mathrm{EO}}_2(x_D=1) &= \xi_2^{\mathrm{EO}} \cos \left( \omega_D t_D \right),
\end{align}
where $\xi_1^{\mathrm{SP}} = -H / \Delta$, $\xi_2^{\mathrm{SP}} = G /\Delta$,  $\xi_1^{\mathrm{EO}} = -K_D /\Delta$, and $\xi_2^{\mathrm{EO}}=K_D /\Delta$.  The homogeneous boundary conditions are unchanged.

In the physical problem, only one of the $p_D$ or $\psi_D$ boundary conditions are inhomogeneous at a time, in either the SP or EO configurations \eqref{eq:periodic-BC-ND}. In the intermediate uncoupled problem, both have opposite sign inhomogeneous driving terms \eqref{eq:periodic-BC-UC}.

\subsection{Oscillatory Solution}
We derive an oscillatory steady-state solution, starting with a transient diffusion problem decomposed as  
\begin{equation}
  \label{eq:periodic-PDE1}
  \delta_i(x_D,t_D) = \delta_{i\mathrm{SS}}(x_D,t_D) + \delta_{i\mathrm{TR}}(x_D,t_D),
\end{equation}
namely a transient exponential decay from an arbitrary initial condition and a steady-state solution with the same frequency as the boundary condition. Here we only seek $\delta_{i\mathrm{SS}}$, 
\begin{equation}
  \label{eq:periodic-PDE2}
  \delta_{i\mathrm{SS}}(x_D,t_D)=\Re\left[ U(x_D) e^{j \omega_D t_D}\right],
\end{equation}
where $j=\sqrt{-1}$, $\Re$ is the real part, and $U(x_D)$ is the complex amplitude. The real part ($|U(x_D)|$) is amplitude and the imaginary part ($\arg\left[ U(x_D)\right]$) is phase shift. First substituting \eqref{eq:periodic-PDE2} into the uncoupled governing equations \eqref{eq:periodic-PDE-UC} and boundary conditions \eqref{eq:periodic-BC-UC}, then equating real and imaginary parts leads to the following second-order ordinary differential equation for the spatial component of the harmonic steady-state behavior
\begin{align}
  \label{eq:periodic-ODE1}
  \frac{\mathrm{d}^2 U}{\mathrm{d} x_D^2} - \frac{j \omega_D}{\Lambda_{ii}} U &= 0,  &
  \left. \frac{\mathrm{d} U}{\mathrm{d} x_D}\right|_{x_D=0}&=0&  U(x_D=1) &= \xi_i^{\{\mathrm{SP,EO}\}}. 
\end{align}
The solution to this ordinary differential equation and boundary conditions is the exponential pair  
\begin{equation}
  \label{eq:periodic-ODE2}
  U_i(x_D) = c_1 e^{\zeta_i x_D} + c_2 e^{-\zeta_i x_D},
\end{equation}
where $\zeta_i=\sqrt{j \omega_D/\Lambda_{ii}}$ and $(c_1,c_2)$ are determined from the applied boundary conditions. Substituting \eqref{eq:periodic-ODE2} into the boundary conditions \eqref{eq:periodic-ODE1} and solving for $(c_1,c_2)$ results in the steady-state solution
\begin{align}
  \label{eq:periodic-soln}
  \delta^{\{\mathrm{SP,EO}\}}_{i\mathrm{SS}}(x_D,t_D)=\Re\left\lbrace \xi^{\{\mathrm{SP,EO}\}}_i e^{j \omega_D t_D } \cosh \left( x_D \zeta_i \right) \mathrm{sech} \left( \zeta_i \right) \right \rbrace.
\end{align}

\end{document}